\documentclass{AIAA}
\usepackage{CJKutf8}
\usepackage{amsmath}
\usepackage{subfig}
\usepackage{overpic}
\usepackage{multirow}
\usepackage{booktabs}
\usepackage{float}
\usepackage[usenames,dvipsnames]{color}
\definecolor{tab_back}{gray}{1}

\usepackage{tikz}
\usepackage{array}
\usepackage{caption}
\usepackage{ulem}
\usepackage{soul}
\usepackage{paralist}
\usepackage{rotating}
\begin{document}
\title{General Reynolds Analogy for Blunt-nosed Bodies in Hypersonic Flows}

\author{Xing-Xing Chen\footnote[2]{Ph. D candidate, School of Physics, 19(A) Yuquan Road.}}
\author{Zhi-Hui Wang\footnote[3]{Associate Prof., School of Physics, 19(A) Yuquan Road.}}
\author{Yong-Liang Yu\footnote[4]{Associate Prof., School of Physics, 19(A) Yuquan Road, email : ylyu@ucas.ac.cn.}}
\affiliation{
University of the Chinese Academy of Sciences,\\
100049 Beijing, The People{'}s Republic of China
}

\begin{abstract}
In this paper, the relation between skin friction and heat transfer along windward sides of blunt-nosed bodies in hypersonic flows is investigated. The self-similar boundary layer analysis is accepted to figure out the distribution of the ratio of skin friction to heat transfer coefficients along the wall. It is theoretically obtained that the ratio depends linearly on the local slope angle of the wall surface, and an explicit analogy expression is presented for circular cylinders, although the linear distribution is also found for other nose shapes and even in gas flows with chemical reactions. Furthermore, based on the theoretical modelling of the second order shear and heat transfer terms in Burnett equations, a modified analogy is derived in the near continuum regime by considering the rarefied gas effects. And a bridge function is also constructed to describe the nonlinear analogy in the transition flow regime. At last, the direct simulation Monte Carlo method is used to validate the theoretical results. The general analogy, beyond the classical Reynolds analogy, is applicable to both flat plates and blunt-nosed bodies, in either continuous or rarefied hypersonic flows.
\\
\\
\\
\\
\\
\\
\\
\\
\\
\end{abstract}

\maketitle
\section*{Nomenclature}
\noindent
\begin{tabular}{@{}lcl@{}}
& &\\
\textit{$C_{f}$} &=& Skin friction coefficient $2\tau_w/\rho_{\infty}U_{\infty}^2$\\
\textit{$C_r$} &=& Proportional factor of the analogy relation, Eq. (\ref{eq_linear_the})\\
\textit{$C_h$} &=& Heat transfer coefficient $2\dot{q}_w/\rho_{\infty}U_{\infty}^3$\\
\textit{$f'$} &=& Non-dimensional velocity along $x$ axis $u/u_e$\\
\textit{$g$} &=& Non-dimensional total enthalpy $H/H_e$\\
\textit{$H$} &=& Total enthalpy per unit mass $(J/kg)$\\
\textit{$k$} &=& Thermal conductivity $(W/(m\cdot K))$\\
\textit{$M_{\infty}$} &=& Mach number in free stream\\
\textit{$n$} &=& Index of the power-law shape leading edge\\
\textit{$Pr$} &=& Prandtl number\\
\textit{$p$} &=& Wall pressure $(N/m^2)$\\
\textit{$\tilde{p}$} &=& Normalized pressure distribution function, Eq. (\ref{eq_distri}) \\
\textit{$\hat{p}_0$} &=& Pressure gradient parameter at the stagnation point $\left[-\frac{d^2(p/p_0)}{d\theta^2}\right]_{\theta=0}$\\
\textit{$\dot{q}_w$} &=& Heat flux to the wall $(W/m^2)$\\
\textit{$r_0$} &=& Section radius of the body of revolution, Eq. (\ref{eq_transform}) $(m)$\\
\textit{$R$} &=& Curvature radius of the surface, Fig. (\ref{fig_sketch}) $(m)$\\
\textit{$R_c$} &=& Radius of the inscribed circle of the base of the body, Fig. (\ref{fig_sketch}) $(m)$\\
\textit{$Re_\infty$}  &=& Free stream Reynolds number $\rho_{\infty}U_{\infty}R_c/\mu_{\infty}$\\
\textit{$Re_l$}  &=& Local Reynolds number $\rho_{\infty}U_{\infty}R/\mu_{\infty}$\\
\textit{$T$} &=& Temperature $(K)$ \\
\textit{$T_0$} &=& Stagnation temperature in free stream $(K)$\\
\textit{$U_{\infty}$} &=& Velocity in free stream $(m/s)$\\
\textit{$u, v$} &=&  Flow velocity along $x$ and $y$ $(m/s)$\\
\textit{$W_r$} &=& Rarefied flow criterion $M_{\infty}^{2\omega}/Re_l$\\
\textit{$x, y$} &=&  Local coordinate in physical space, Fig. (\ref{fig_sketch}) $(m)$\\
\textit{$\beta$} &=& Gradient parameter $2 d (\ln u_e)/d (\ln \xi)$\\
\textit{$\gamma$} &=& Specific heat ratio\\
\textit{$\Gamma_{1,2}$} &=& Coefficients in Eq. (\ref{eq_Gamma}) \\
\textit{$\eta, \xi$} &=& Local coordinate in $Lees-Dorodnitsyn$ transformation space \\
\textit{$\theta$} &=& Slope angle of the surface, Fig. (\ref{fig_sketch})\\
\textit{$\theta_c$} &=& Surface slope angle at the base of the body, Fig. (\ref{fig_sketch})\\
\textit{$\mu$} &=& Viscosity $(kg/(m\cdot s))$\\
\textit{$\rho$} &=& Density $(kg/m^3)$\\
\textit{$\varsigma$} &=& Normalized heat transfer distribution function, Eq. (\ref{eq_distri}) \\
\textit{$\tau_w$} &=& Skin shear stress ($N/M^2$)\\
\textit{$\omega$} &=& Index of the viscosity-temperature power law \\
& &\\
Subscripts& &\\
\textit{$_0$} &=& Stagnation point condition\\
\textit{$_e$} &=& Boundary layer edge condition\\
\textit{$_{free}$} &=& Free molecular flow condition\\
\textit{$_{r}$} &=& Reference state\\
\textit{$_w$} &=& Wall condition\\
\textit{$_{\infty}$} &=& Free stream condition\\
& &\\
Superscripts& &\\
\textit{$^{(1)}$} &=& First order approximation based on Navier-Storks-Fourier equations\\
\textit{$^{(2)}$} &=& Second order correction based on Burnett equations\\
\end{tabular}

\section{Introduction}

There have long been concerns on the connection between skin friction and wall heat-transfer rate (or simply, heat transfer), as macroscopically the two quantities are related respectively to the normal gradient of velocity and temperature, and microscopically they are momentum and energy transports arising from molecular moves and collisions.
The most famous result of this subject is the Reynolds analogy in the flat plate boundary layer flow problem, where the skin friction and heat transfer were found proportional along the surface.
But such simple relationship does not exist for curved surfaces. For example, in flows past circular cylinders or spheres the heat transfer reaches its maximum at the stagnation point and diminishes monotonously downstream while the skin friction is zero at the stagnation point and varies non-monotonically downstream. With regard to the relation between skin friction and heat transfer, there has not yet been a theory suitable for the curved surfaces in either continuum or rarefied gas flows.

Separately speaking, the heat transfer in hypersonic flows has received much more attentions than the skin friction due to early engineering requirements. Lees \cite{lees1956} first developed theories of the heat transfer to blunt-nosed bodies in hypersonic flows based on laminar boundary layer equations. The method made rational predictions in high Reynolds number flows. Later Fay and Riddell \cite{fay1958} gave more detailed discussions on the stagnation point heat transfer and developed prediction formulas with improved accuracy. Considering the downstream region of the stagnation point, Kemp et al. \cite{kemp1959} made an improvement to extend the theory to more general conditions. In practice, empirical formulas of the heat transfer distribution were also constructed for typical nose shapes such as circular cylinders and spheres \cite{murzinov1966,beckwith1961}.

Besides above boundary layer analyses in continuum flows, theoretical studies of heat transfer in rarefied gas flows have been carried out by different approaches. Cheng \cite{cheng1961} accepted the thin viscous shock layer equations and obtained analytical expressions of the stagnation point heat transfer from the boundary layer flow to the free molecular flow. Wang et al. \cite{wang2009,wang2010} presented a theoretical modelling of the non-Fourier aeroheating performance at the stagnation point based on the Burnett equations. A control parameter $W_r$ was derived as the criterion of the local rarefied gas effects, and formulas based on the parameter were found to correlate the heat transfer both at the stagnation point and in the downstream region.

Unlike the heat transfer, the skin friction is often neglected in continuum flows over large blunt bodies, for the friction drag is much less than the pressure drag in those conditions. Most of the existing studies concerns only about turbulence flows rather than laminar flows.
However for wedges or cones with small angles in rarefied gas flows, the skin friction contributes a significant part, as much as $50\%$, in the total drag \cite{santos2002}. Unfortunately, there is still no reliable theory we could use to predict the skin friction over curved surfaces. It will be meaningful if we find a general Reynolds analogy by using which we could estimate the skin friction based on available heat transfer prediction formulas, or vice versa. In Lees and the followers' research on the aeroheating performance of blunt bodies, the momentum equations were solved coupled with the energy equation of boundary layer flows, which offers a breakthrough point to analyze the relation between the skin friction and the heat transfer.

In the present work, the ratio of skin friction to heat transfer along curved surfaces is firstly discussed based on the self-similar solution of boundary layer equations. An expression with simple form is obtained for circular cylinders as a typical example. Subsequently, an extended analogy is deduced in the near continuum flow regime by considering the non-linear shear and heat transfer in the Burnett approximation, and it is found the rarefied gas effects on the analogy are characterized by the rarefied flow criterion introduced in our previous study. As a preliminary study, the molecular vibration and chemical reaction effects are out of consideration in the theoretical analysis. The direct simulation Monte Carlo (DSMC) method \cite{bird1994} is also used to simulate present flows to validate the theoretical results.

\section{Ratio of Skin Friction to Heat Transfer in Continuous Flows}
\subsection{Linear Analogy for Blunt-nosed Bodies.}
\begin{figure}[htbp]
\begin{overpic}[width=250pt]{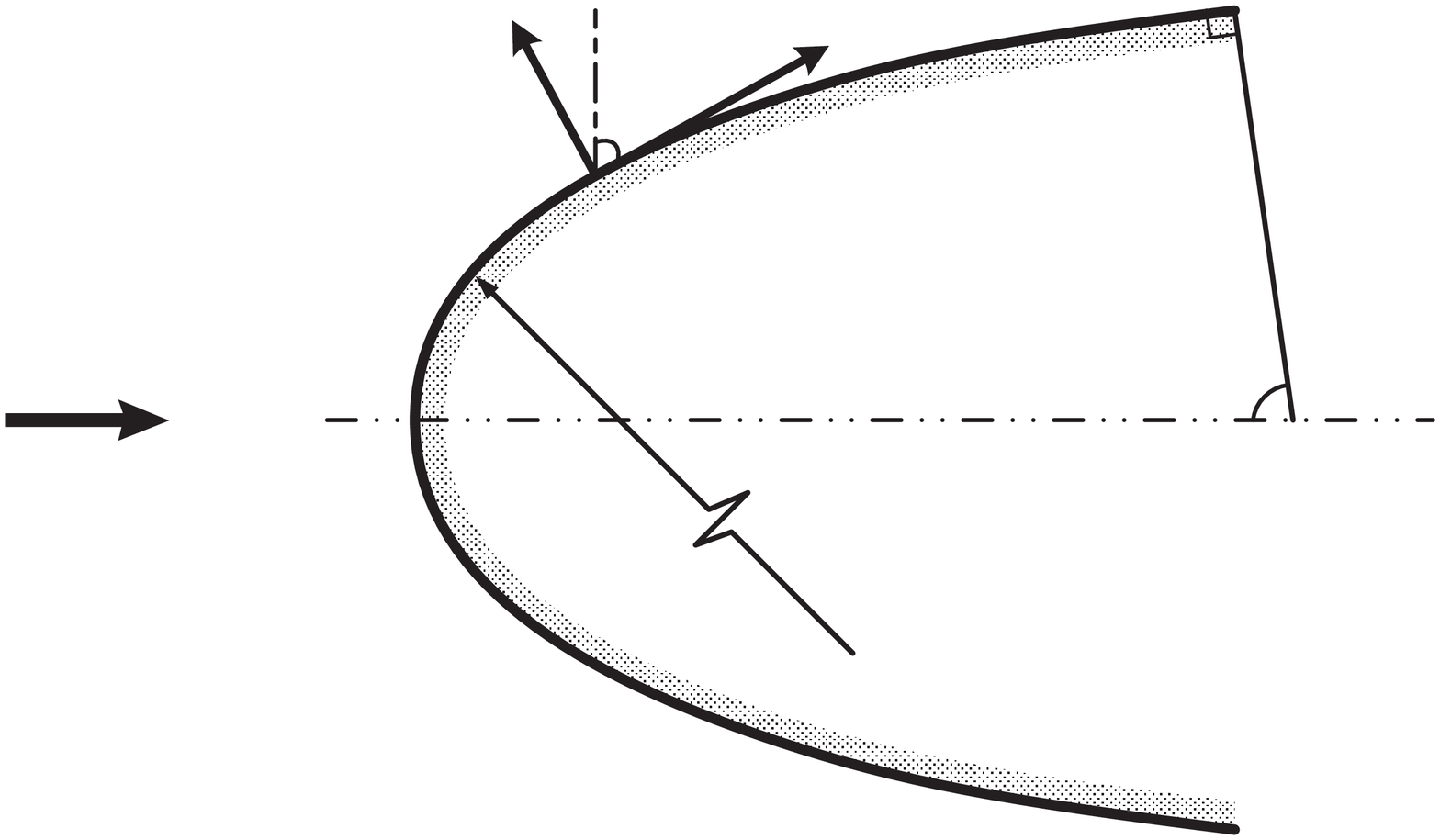}
\small
\put(35,58){$y$}
\put(57,56){$x$}
\put(0,33){$M_{\infty}\gg1$}
\put(44,51){$\theta$}
\put(83,33){$\theta_c$}
\put(81,42){$R_c$}
\put(41,22){$R$}
\end{overpic}
\caption{Blunt-nosed body in a hypersonic flow.}\label{fig_sketch}
\end{figure}

Fig. \ref{fig_sketch} is a sketch illustrating the hypersonic flow over a blunt-nosed cylindrical body or body of revolution. The local coordinate is set on the wall.

In order to seek the self-similar solutions of the boundary layer equations governing the flow around blunt-nosed bodies, Lees et al. \cite{lees1956} introduced the $Lees-Dorodnitsyn$ coordinate transformation:
\begin{equation}
	\begin{split}\label{eq_transform}
	\xi&=\int_{0}^{x} \rho_e u_e \mu_e r_0^{2m}dx
	\\
	\eta&=\frac{u_e}{\sqrt{2\xi}}\int_0^y r_0^m \rho dy
	\end{split}
\end{equation}
and the normalizations of velocity and temperature:
\begin{equation}
	\begin{split}\label{eq_fg}
	f'(\eta)&=u/u_e\\
	g(\eta)&=H/H_e.
	\end{split}
\end{equation}
where $m=0$ for planar bodies and $m=1$ for bodies of revolution. With the transformation, the boundary layer equations were simplified and the self-similar solution have been obtained under certain conditions \cite{anderson2006,lees1956,fay1958}.

Defining the coefficients of skin friction and heat transfer as $C_f=2\tau_w/\rho_{\infty}U_{\infty}^2$ and \break $C_h=2 \dot{q}_w/\rho_{\infty}U_{\infty}^3$, respectively, and with the approximation $H_e\approx\frac{1}{2}U_{\infty}^2$ in the hypersonic limit ($M_{\infty}\gg1$), we have:
\begin{equation}
	\begin{split}\label{eq_ratio}
	\frac{C_f}{C_h}=2Pr\frac{u_ef''_w}{U_{\infty}g'_w}
	\end{split}
\end{equation}
from Eqs. (\ref{eq_transform}) and (\ref{eq_fg}).

From the compressible Bernoulli's equation, the streamwise velocity $u_e$ along the edge of the boundary layer depends mainly on the wall pressure $p$ \cite{lees1955}:
\begin{equation}\label{eq_stream_velo}
	\frac{u_e}{U_{\infty}}=\sqrt{\left(1+\frac{2}{\gamma-1}\frac{1}{M_{\infty}^2}\right)\left[1-\left(\frac{p}{p_0}\right)^{(\gamma-1)/\gamma}\right]}
\end{equation}
On the basis of the symmetry and smoothness of the pressure distribution near the stagnation point, Eq.~(\ref{eq_stream_velo}) can be linearized to:
\begin{equation}\label{eq_velTotheta}
	\frac{u_e}{U_{\infty}}\approx\sqrt{\hat{p}_0 \frac{\gamma-1}{2\gamma}}\cdot\theta
\end{equation}
with $M_\infty\gg1$ and $\theta\ll1$, where $\hat{p}_0=\left[-\frac{d^2(p/p_0)}{d\theta^2}\right]_{\theta=0}$.
In fact it was found that the linear variation of $u_e$ with $\theta$ is valid downstream the stagnation point till $\theta=4\pi/9$ from Korobkin's experimental data \cite{lees1955,korobkin1954}. Therefore in following analyses, Eq. (\ref{eq_velTotheta}) is accepted not only near the stagnation point, but also in the downstream region.

Cohen and Reshotko \cite{cohen1956} numerically solved the self-similar boundary layer equations and calculated $2f''_w/g'_w$. The value depends on $T_w/T_0$ and varies with $\beta=2 d (\ln u_e)/d (\ln \xi)$ along the surface. Actually according to the calculations from Kemp et. al \cite{kemp1959}, $g'_w$ can be taken as a constant along the isothermal wall with a slowly changing curvature radius. And from Cohen and Reshotko's solutions \cite{cohen1956}, the slight variation of $f''_w$ is of the same order as that of $g'_w$, and thus it is reasonable to assume a constant $2f''_w/g'_w$ as long as the self-similar assumption is satisfied.

As a result, the expression of $C_f/C_h$ in Eq. (\ref{eq_ratio}) can be written as:
\begin{equation}\label{eq_linear_the}
	\frac{C_f}{C_h}=C_r\theta
\end{equation}
with the coefficient
\begin{equation*}
	C_r=2Pr\frac{f''_w}{g'_w}\sqrt{\hat{p}_0 \frac{\gamma-1}{2\gamma}}
\end{equation*}
being independent of the location. The equation indicates that for a variety of nose shapes in hypersonic flows, $C_f/C_h$ is proportional to $\theta$ along the windward surface as long as $T_w$ is a constant. Thus, we have a more general form of the analogy relation between skin friction and heat transfer which is not restricted to the flat plate flow, compared with the classical Reynolds analogy.

\subsection{Analogy for Circular Cylinders.}
In order to further calculate the coefficient $C_r$, the shape dependent $f''_w/g'_w$ and $\hat{p}_0$ should be specified. As a typical example, the corresponding values for circular cylinders are to be presented in the following.

First, the pressure distribution can be calculated by the Newtonian-Busemann theory \cite{anderson2006}, and at the stagnation point, we have
\begin{equation}\label{eq_p_grad}
	\hat{p}_0=3
\end{equation}

\begin{figure}[htbp]
	\centering
	\begin{minipage}[]{0.5\linewidth}
	\centering
	\begin{overpic}[width=200pt]{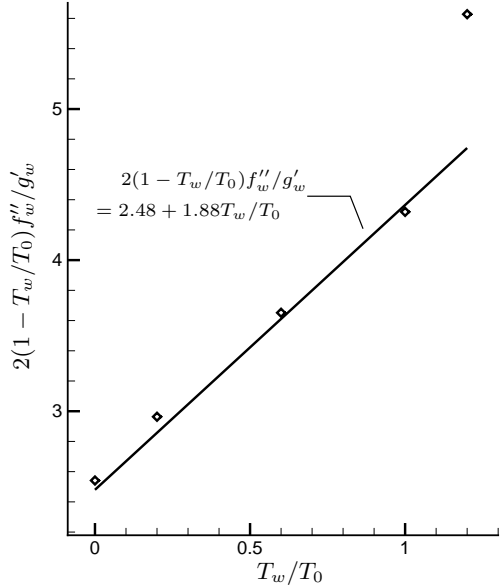}
	\small
	\put(2,35){\begin{turn}{90}$2(1-T_w/T_0)f''_w/g'_w$\end{turn}}
	\put(42,1){$T_w/T_0 $}
	\scriptsize
	\put(20,65){$2(1-T_w/T_0)f''_w/g'_w$}
	\put(16,60){$=2.48+1.88T_w/T_0$}
	\end{overpic}
	\end{minipage}
	\caption{Linear correlation of $2(1-T_w/T_0)f''_w/g'_w$ at the stagnation point of two-dimensional ($\beta=1$) bodies against $T_w/T_0$ based on data in \cite{cohen1956}.}\label{fig_corre_fg}
\end{figure}

Second, since $2f''_w/g'_w$ is regarded as a constant along the surface, we will calculate the value at the stagnation point where $\beta=1$ for a planar body. Then from Cohen and Reshotko's numerical calculations \cite{cohen1956}, $2f''_w/g'_w$ was obtained within $0\le T_w/T_0\le1.2$ as plotted in Fig. (\ref{fig_corre_fg}). Obviously the points with $T_w/T_0\le1$ fall into a straight line, and a linear correlation
\begin{equation}\label{eq_corre_fg}
	2(1-T_w/T_0)f''_w/g'_w=2.48+1.88T_w/T_0
\end{equation}
fits well with the points.
With Eqs.~(\ref{eq_p_grad}) and (\ref{eq_corre_fg}) applied to Eq.~(\ref{eq_linear_the}), we have
\begin{equation}\label{eq_ratio2}
	C_f/C_h=Pr\sqrt{\frac{3(\gamma-1)}{2\gamma}}\frac{2.48+1.88T_w/T_0}{1-T_w/T_0}\theta
\end{equation}

The explicit relation in Eq. (\ref{eq_ratio2}) gives a convenient formula to predict $C_f/C_h$ around a circular cylinder in hypersonic flows. Comparisons are given in the following with DSMC numerical simulations of the nitrogen gas flows past blunt-nosed bodies in hypersonic speeds under $M_\infty=5\sim25$, $Re_\infty=300\sim1500$, $T_w/T_0=0.01\sim0.4$. The simulations are carried out by using the source code as described in \cite{wang2010}. The molecular vibration and chemical reaction effects are excluded to correspond with the derivations in this paper.

\pgfsetxvec{\pgfpoint{1pt}{0}}
\pgfsetyvec{\pgfpoint{0}{1pt}}
\begin{figure}[htbp]
	\centering
 \begin{minipage}[t]{0.5\linewidth}
 	\centering
	\begin{overpic}[width=200pt]{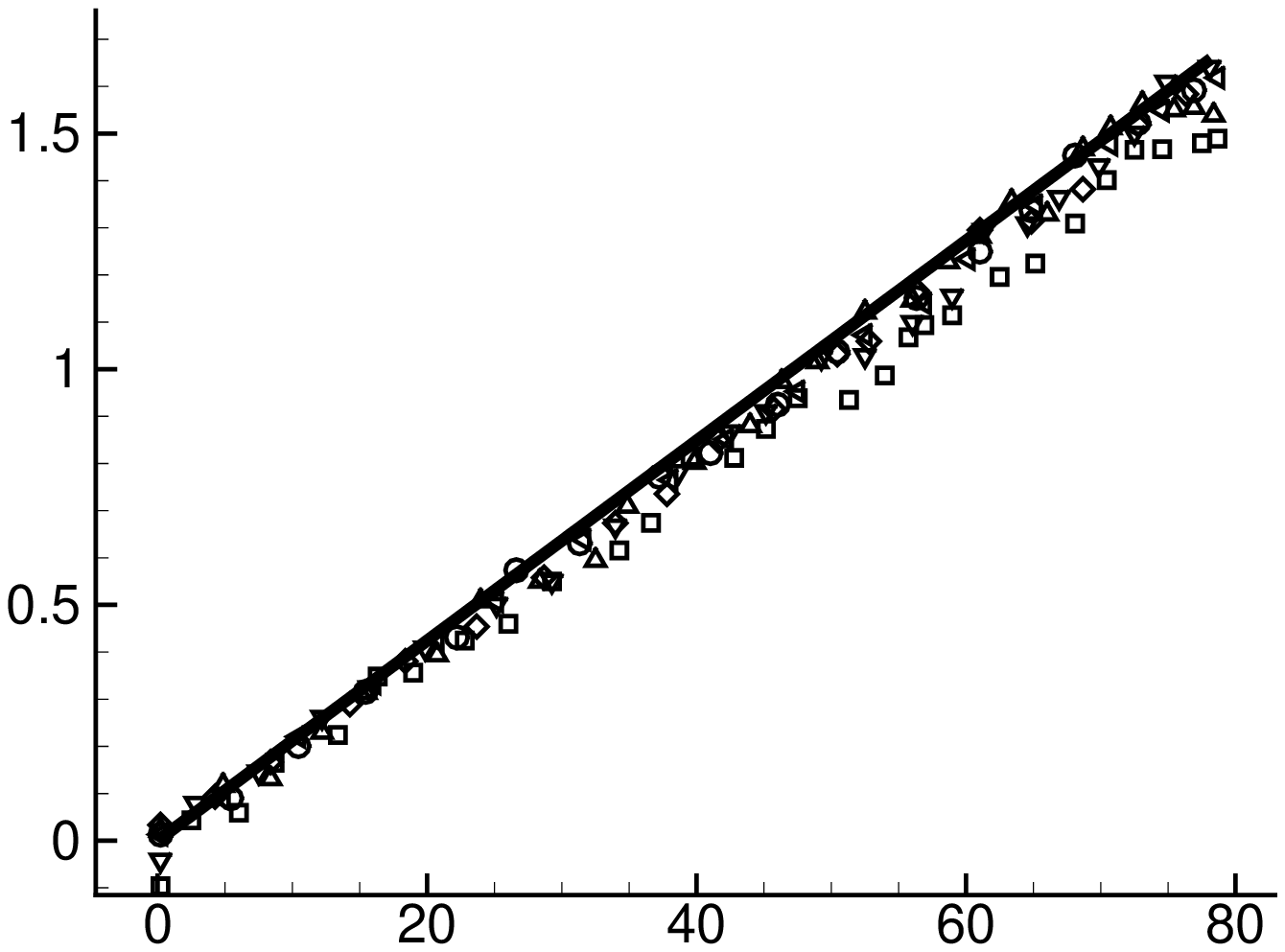}
	\put(20,50){\includegraphics[width=25pt,grid]{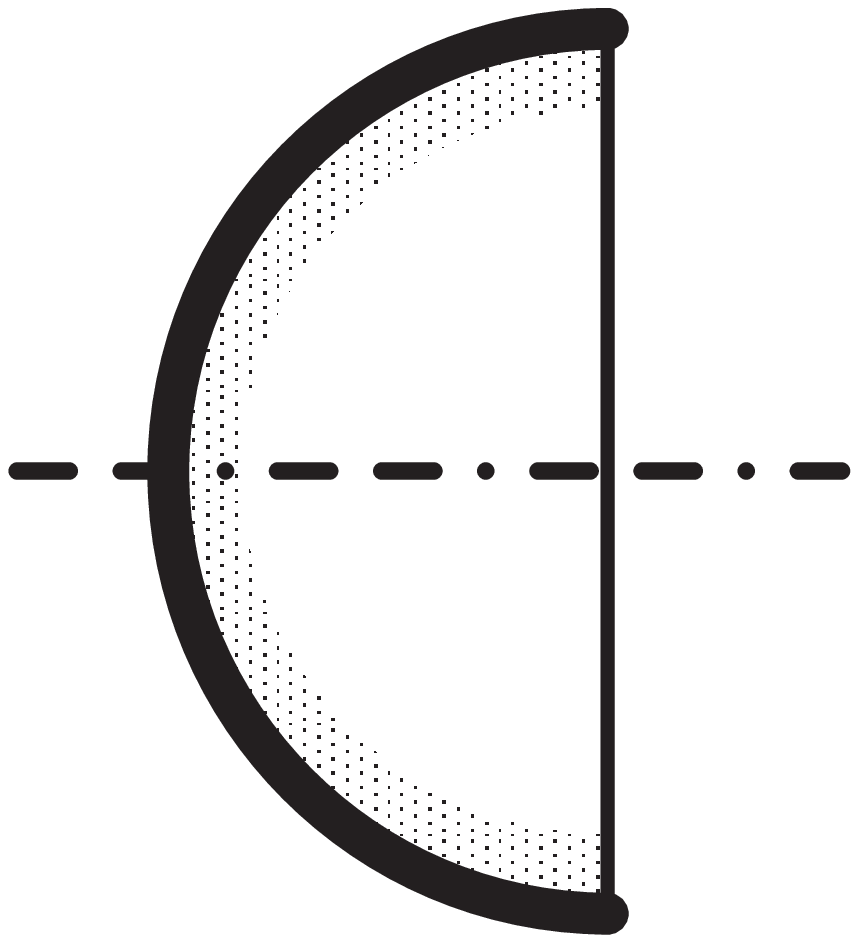}}
	\put(53,0){$\tfrac{180}{\pi}\theta(^{\circ})$}
	\large
	\put(0,38){$\frac{C_f}{C_h}$}
	\scriptsize
	\put(45,25){\tikz \draw[line width=1.0pt,scale=0.7] (0,25)--(20,25);: Eq. (\ref{eq_ratio2})}
	\put(38,15){Symbols: DSMC Simulations}
	\put(20,65){Circular Cylinder}
	\put(5,-10){
	\small
	\begin{tabular}
	{cp{25pt}<{\centering}p{20pt}<{\centering}p{20pt}<{\centering}p{20pt}<{\centering}p{20pt}<{\centering}p{20pt}<{\centering}}
	Labels&
      \tikz \draw[line width=0.7pt,scale=0.7] (0,0) rectangle (3.5,3.5);
      &
      \tikz \draw[line width=0.7pt,scale=0.7] (0,0)--(2,4)--(4,0)--cycle;
      &
      \tikz \draw[line width=0.7pt, rotate=180,,scale=0.7] (0,0)--(2,4)--(4,0)--cycle;
      &
      \tikz \draw[line width=0.7pt,rotate=45,scale=0.7] (0,0) rectangle (3.5,3.5);
      &
      \tikz \draw[line width=0.7pt, rotate=90,scale=0.7] (0,0)--(2,4)--(4,0)--cycle;
      &
      \tikz \draw[line width=0.7pt,scale=0.7] (2,2) circle(2);
      \\
      \hline
      $M_{\infty}$&5&8&10&15&20&25\\
	\end{tabular}
	}
	\end{overpic}
  \vspace{30pt}
 \end{minipage}
 \centering
 \caption{Distributions of $C_f/C_h$ for circular cylinders under different Mach numbers with $T_w/T_0=0.03$.\label{fig_mach}}
\end{figure}

\begin{figure}[htbp]
	\centering
	\begin{minipage}[t]{0.5\linewidth}
	\centering
	\begin{overpic}[width=200pt]{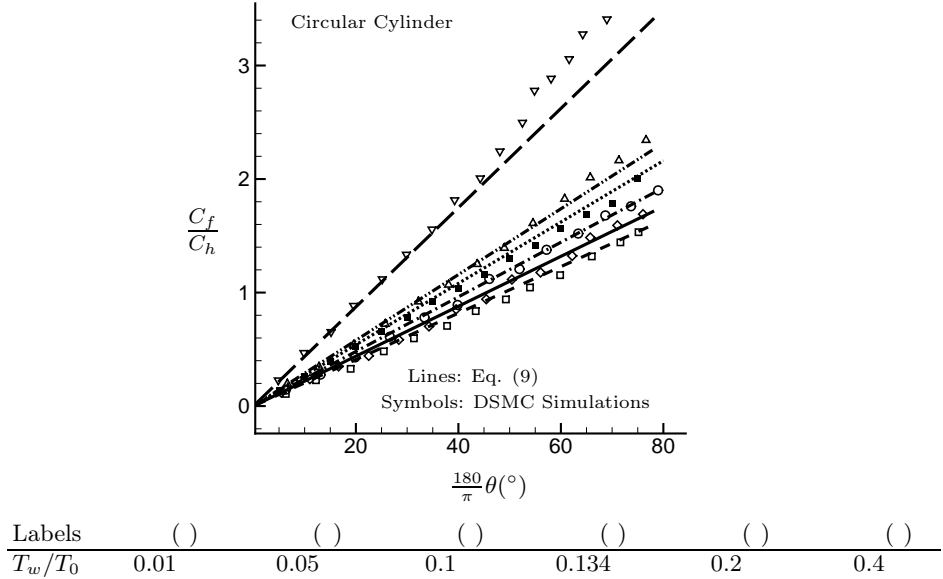}
	\put(50,2){$\tfrac{180}{\pi}\theta(^{\circ})$}
	\large
	\put(0,50){$\frac{C_f}{C_h}$}
	\scriptsize
	\put(42,23){Lines: Eq. (\ref{eq_ratio2})}
	\put(37,18){Symbols: DSMC Simulations}
	\put(20,90){Circular Cylinder}
	\put(-35,-10){
	\small
	\begin{tabular}
	{p{25pt}<{\centering}p{50pt}<{\centering}p{50pt}<{\centering}p{50pt}<{\centering}p{50pt}<{\centering}p{50pt}<{\centering}p{50pt}<{\centering}}
	Labels&
      \tikz[baseline=-2pt,dash pattern=on 4.5 off 4.5]
      \draw[line width=1.3] (0,0)--(18,0);
      (\tikz[baseline=-1pt] \draw[line width=0.7pt,scale=0.7] (0,0) rectangle (3.5,3.5);)
      &
      \tikz[baseline=-2pt]
      \draw[line width=1.3] (0,0)--(18,0);
      (\tikz \draw[line width=0.7pt,rotate=45,scale=0.7] (0,0) rectangle (3.5,3.5);)
      &
      \tikz[baseline=-2pt,dash pattern=on 3 off 1.3 on 0.8 off 1.3]
      \draw[line width=1.3] (0,0)--(18,0);
      (\tikz[baseline=-1pt] \draw[line width=0.7pt,scale=0.7] (2,2) circle(2);)
      &
      \tikz[baseline=-2pt,dash pattern=on 1.2 off 1.2]
      \draw[line width=1.3] (0,0)--(18,0);
      (\tikz[baseline=-1pt] \draw[line width=0.7pt,scale=0.7,fill] (0,0) rectangle (3.5,3.5);)
      &
      \tikz[baseline=-2pt,dash pattern=on 8 off 1.3 on 0.8 off 1.3 on 0.8 off 1.3]
      \draw[line width=1.3] (0,0)--(18,0);
      (\tikz[baseline=-1pt] \draw[line width=0.7pt,scale=0.7] (0,0)--(2,4)--(4,0)--cycle;)
      &
      \tikz[baseline=-2pt,dash pattern=on 10 off 3]
      \draw[line width=1.3] (0,0)--(18,0);
      (\tikz[baseline=-1pt] \draw[line width=0.7pt,scale=0.7] (2,0)--(0,4)--(4,4)--cycle;)
      \\
      \hline
      $T_w/T_0$&$0.01$&$0.05$&$0.1$&$0.134$&$0.2$&$0.4$\\
	\end{tabular}
	}
	\end{overpic}
    \vspace{30pt}
	\end{minipage}
	\caption{Distributions of $C_f/C_h$ for circular cylinders in different wall temperatures with $M_{\infty}=10$, except that $T_w/T_0=0.134$ case is from Santos \cite{santos2002} under $M_\infty=12$ and $Re_\infty=214$.}\label{fig_temp}
\end{figure}

The numerically computed distributions of $C_f/C_h$ along windward surfaces of circular cylinders are presented in Fig. (\ref{fig_mach}) and (\ref{fig_temp}). In Fig. (\ref{fig_mach}), the numerical results from different $M_\infty$ all fit well with Eq.~(\ref{eq_ratio2}), showing a Mach number independence, except that the ratio with $M_\infty=5$, the lower limit of hypersonic flows, is slightly lower. Meantime cases with $T_w/T_0=0.01\sim 0.4$ in Fig. (\ref{fig_temp}), as well as a reference result from Santos \cite{santos2002} with $T_w/T_0=0.134$, indicate that Eq. (\ref{eq_ratio2}) precisely describes the variation of $C_f/C_h$ with $T_w/T_0$. Other flow parameters in simulations are $R_c=2\times10^{-3}m, T_\infty=300K, \rho_\infty=1.5\times10^{-3}kg/m^3$, and $Pr=0.71,~\gamma=1.4$ are accepted in Eq. (\ref{eq_ratio2}) for comparisons.

The Reynolds analogy for the flat plate flow shows that $C_f/C_h$ distributes uniformly along the surface as mentioned. Throughout the zero-thickness leading edge, the ratio can be expressed as:
\begin{equation}\label{eq_ratio_plate}
	\frac{C_f}{C_h}=\left[\frac{1}{2}+\frac{\gamma+1}{2\gamma(\gamma-1)M_{\infty}^2}-\frac{\gamma+1}{4\gamma}\frac{T_w}{T_0} \right]^{-1}
\end{equation}
The equation was derived from the free molecular theory, and with $M_{\infty}\gg1$, $T_w/T_0\ll1$, its range of application could be extended from the free molecular flow to the boundary layer flow \cite{chen2014}.

\begin{figure}[htbp]
	\centering
	\begin{minipage}[]{0.5\linewidth}
	\centering
	\begin{overpic}[width=200pt]{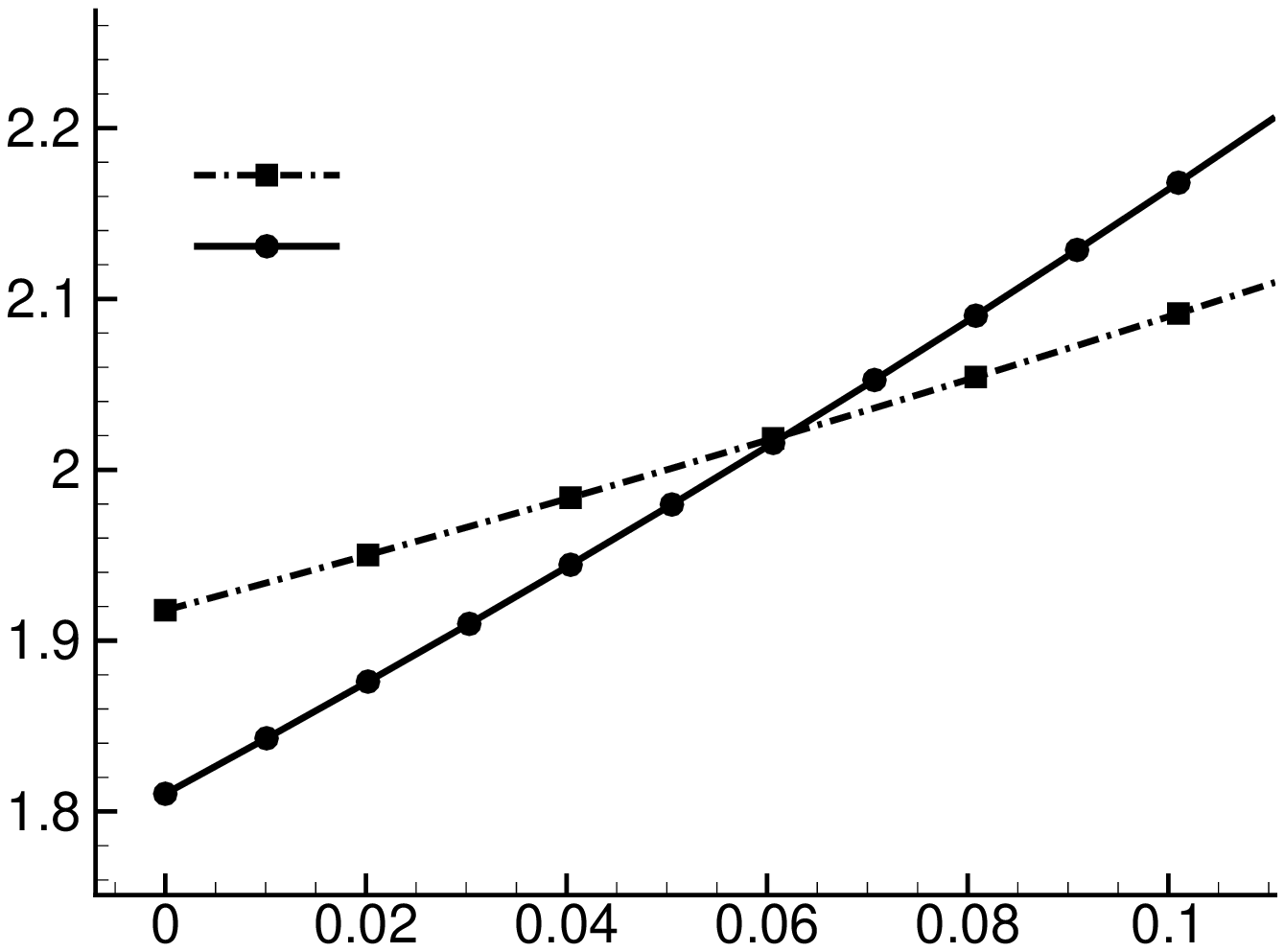}
	\put(2,38){$\frac{C_f}{C_h}$}
	\small
	\put(45,1){$T_w/T_0$}
	\scriptsize
	\put(33,59){Flat plate, Eq. (\ref{eq_ratio_plate})}
	\put(33,54.2){General analogy, Eq. (\ref{eq_ratio_pi})}
	\put(60,20){\includegraphics[width=70pt]{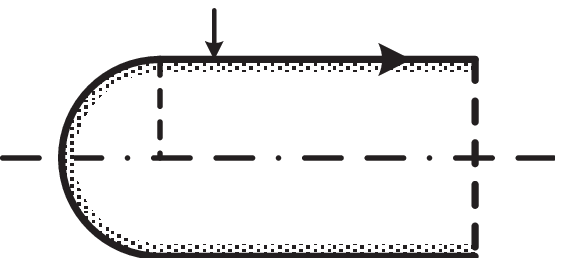}}
	\tiny
	\put(66,38){Flat Segment}
	\put(85,34){$x$}
	\put(71,28){$R_c$}
	\end{overpic}
	\end{minipage}
	\caption{Comparison between the classical Reynolds analogy and the linear analogy for curved surfaces. $M_{\infty}=10$, $\gamma=1.4$ and $Pr=0.71$ in Eq. (\ref{eq_ratio_plate}) and (\ref{eq_ratio_pi}).}\label{fig_rey_ana}
\end{figure}

In practice, a finite thickness and a blunt nose always exist in the leading edge of a flat plate as illustrated in Fig. (\ref{fig_rey_ana}). Assuming a cylindrically blunted nose in the front of the plate, the ratio $C_f/C_h$ at the flat segment can be obtained by taking $\theta={\pi}/{2}$ in Eq. (\ref{eq_ratio2}) as:
\begin{equation}\label{eq_ratio_pi}
	\frac{C_f}{C_h}=\frac{\pi Pr}{2}\sqrt{\frac{3(\gamma-1)}{2\gamma}}\frac{2.48+1.88T_w/T_0}{1-T_w/T_0}
\end{equation}
Values of $C_f/C_h$ in Eq. (\ref{eq_ratio_pi}) and for zero-thickness flat plate are plotted in Fig. (\ref{fig_rey_ana}) against ${T_w}/{T_0}$. The difference between them is less than $5\%$ with $0\le {T_w}/{T_0}\le 0.1$. This result indicates that the present linear analogy in flows around curved bodies is in fact consistent with the classical Reynolds analogy for flat plate flows.

\subsection{Analogy for other nose shapes and flow conditions.}

\begin{figure}[htbp]
\begin{minipage}[]{0.4\linewidth}
\begin{overpic}[width=250pt]{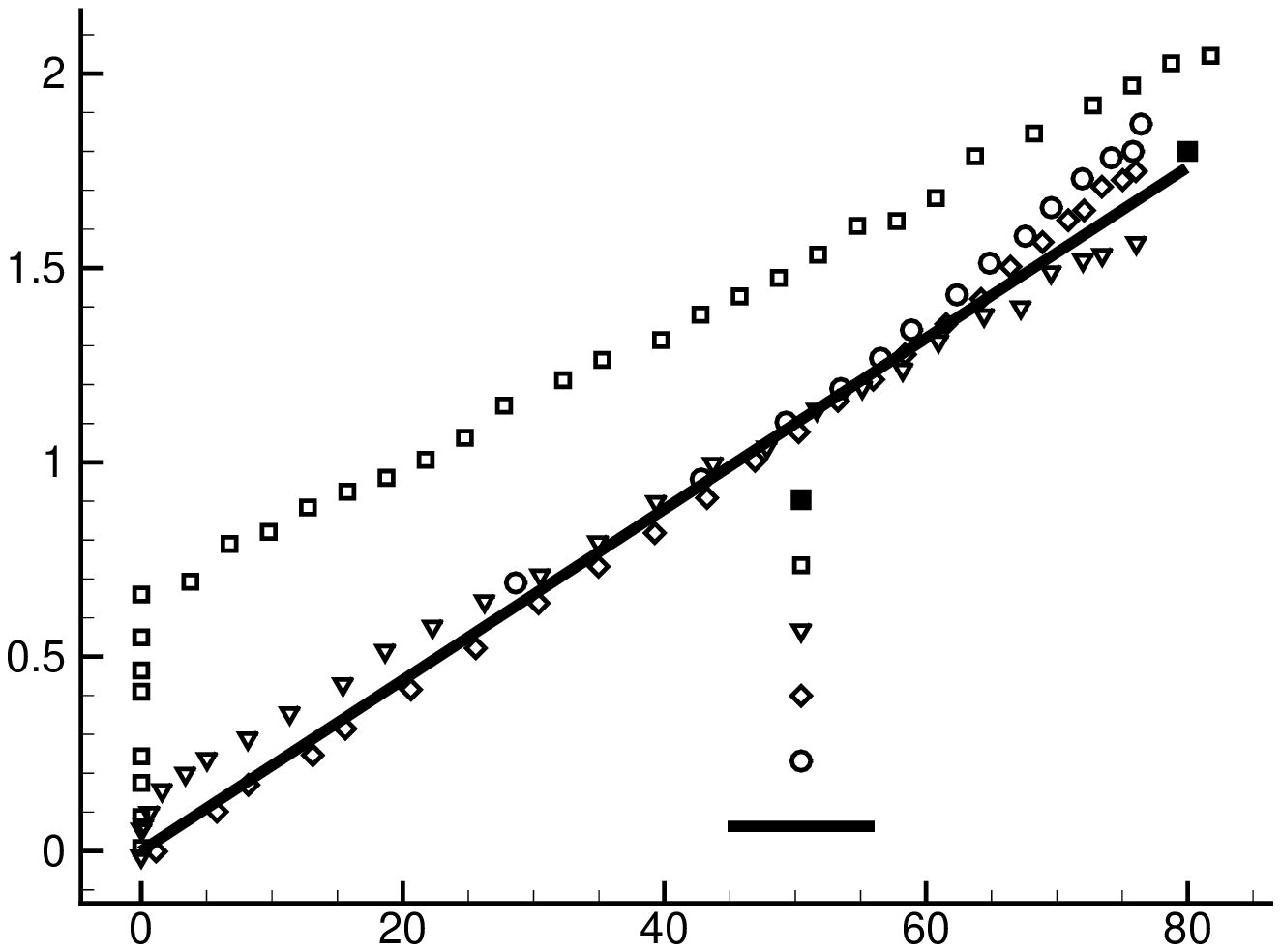}
	\put(50,0){$\tfrac{180}{\pi}\theta(^{\circ})$}
	\large
	\put(3,40){$\frac{C_f}{C_h}$}
	\scriptsize
	\put(72,37){$10^\circ$ wedge}
	\put(72,32.5){Flat Nosed}
	\put(72,28){Power-law $n=0.3$}
	\put(72,23.5){Power-law $n=0.5$}
 	\put(72,19.3){Power-law $n=0.7$}
	\put(72,14.8){Eq. (\ref{eq_ratio2})}
\end{overpic}
\end{minipage}
\hspace{80pt}
{\begin{minipage}[]{0.3\linewidth}
\begin{overpic}[width=120pt]{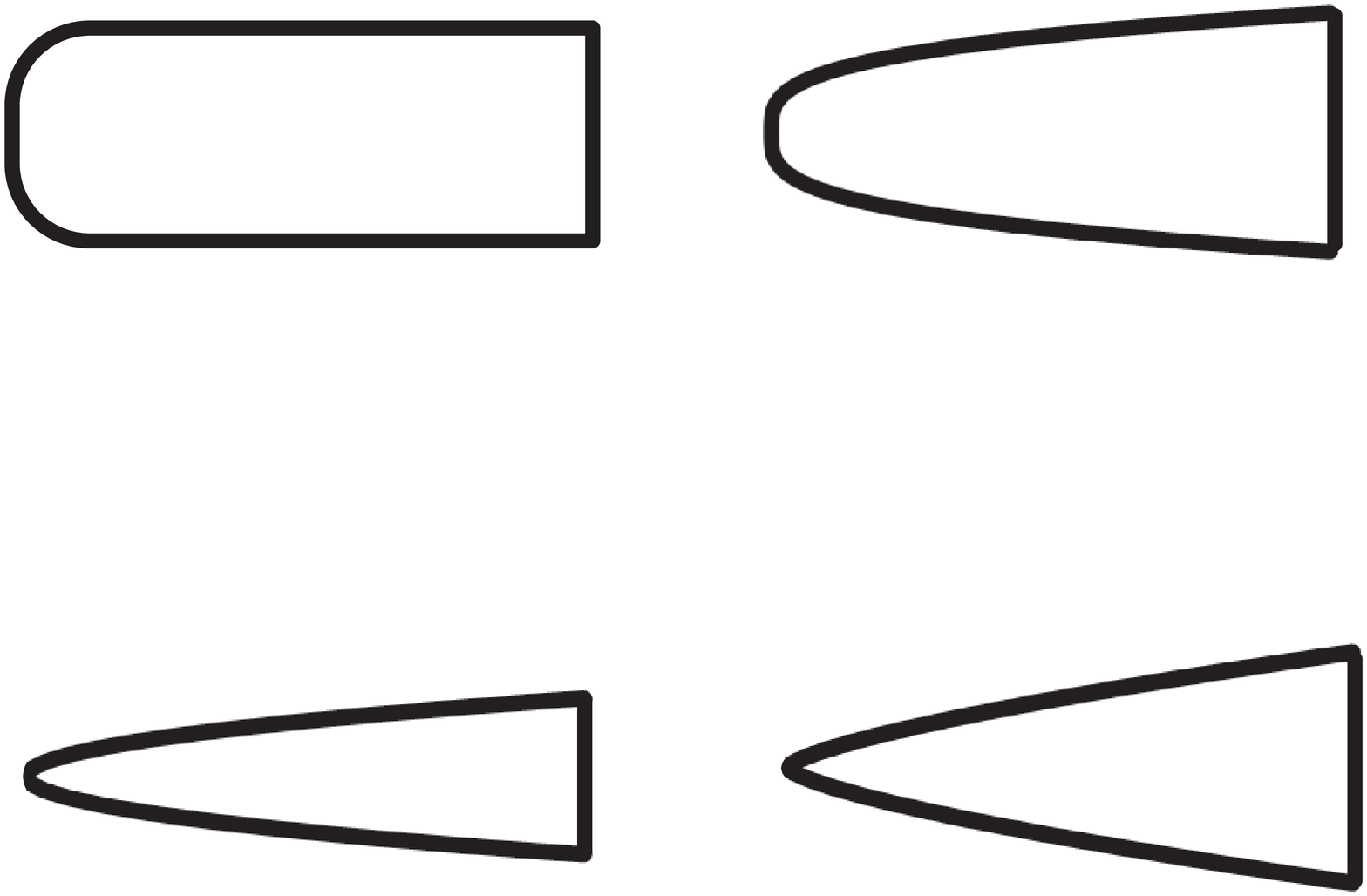}
\scriptsize
\put(2,-20){rectangle and power-law shapes.}
\put(-3,35){rectangle cylinder}
\put(60,35){$n=0.3$}
\put(10,0){$n=0.5$}
\put(60,0){$n=0.7$}
\end{overpic}
\end{minipage}}
\caption{$C_f/C_h$ from DSMC simulations of different shapes of blunt-nosed bodies with $M_{\infty}=10$ , $Re_\infty=200$ and $T_w/T_0=0.05$. }\label{fig_shapes}
\end{figure}

Although lack of explicit expressions at present, due to the mathematical complexity, the existence of the self-similar boundary layer flow could also be found near the wall surfaces of many other nose shapes as long as the variation of the radius of curvature is slow \cite{anderson2006}. Flows over two dimensional wedges with a variety of shapes under $M_\infty=10,~T_w/T_0=0.05$ are also simulated to extend our discussions, as shown in Fig. (\ref{fig_shapes}). Among these results, the shape of the power-law leading edge is expressed as $y=cx^n$, and $\theta_c=80^\circ$ in the present simulations. In simulations $Re_\infty=200, R_c=2\times10^{-4}m, T_\infty=300K$, $\rho_\infty=4\times10^{-3}kg/m^3$. Distributions of $C_f/C_h$ along surfaces of power-law shapes with $n=0.5, 0.7$ are both linear with $\theta$ as expected. The lines diverge slightly from each other but still fit with Eq. (\ref{eq_ratio2}) generally. The discrepancies may be caused mainly by the variation of the wall pressure distributions.

The present theory is not suitable for the rectangle cylinder with a round shoulder, since the boundary layer around the body is highly non-selfsimilar \cite{kemp1959}. As presented in Fig. (\ref{fig_shapes}), $C_f/C_h$ increases in the vertical windward segment with $\theta$ maintaining zero. A similar variation can be observed near the stagnation point of the power-law shape with $n=0.3$, where the surface is also nearly vertical and $C_f/C_h$ increases sharply. However, the linear variation of $C_f/C_h$ with $\theta$ is still observed in the round shoulder segment of the rectangle cylinder and the downstream region of the $n=0.3$ power-law shape. This phenomenon indicates a possibility that the present linear analogy relation could be extended to wider situations.

\begin{figure}[htbp]
\centering
\begin{overpic}[width=200pt]{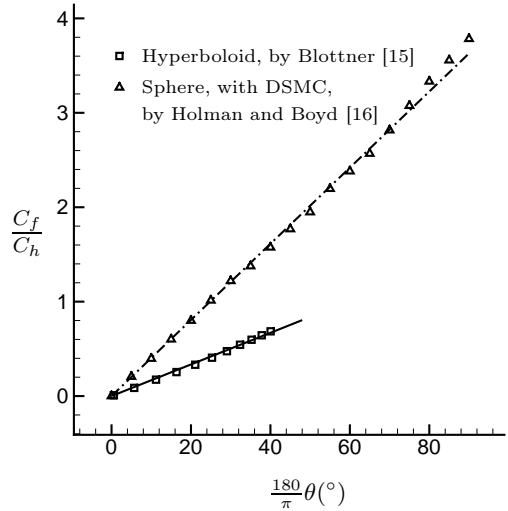}
\put(50,1){$\tfrac{180}{\pi}\theta(^{\circ})$}
\large
\put(1,50){$\frac{C_f}{C_h}$}
\scriptsize
\put(26,84){Hyperboloid, by Blottner \cite{blottner1970}}
\put(26,78){Sphere, with DSMC,}
\put(26,73){by Holman and Boyd \cite{boyd2011}}
\end{overpic}
\caption{$C_f/C_h$ from numerical simulations with real gas effects.\label{fig_real}}
\end{figure}

Further more, $C_f/C_h$ is also found proportional to $\theta$ when considering the real gas effects. Blottner \cite{blottner1970} calculated the boundary layer equations for the equilibrium air flow over a hyperboloid under $M_\infty=20$. Holman and Boyd \cite{boyd2011} computed the dissociating air flow over a sphere under $M_\infty=25$ and $Kn_\infty=0.002$ ($Kn_\infty=\lambda_\infty/R$, where $\lambda_\infty$ is the mean free path of molecules in the free stream) with both DSMC method and Navier-Storks(N-S) equations. Results of $C_f/C_h$ from both hyperboloid and sphere cases are demonstrated in Fig. (\ref{fig_real}). Although the slopes are much different from that of the above-mentioned calorically perfect gas condition, the linear feature can be still clearly observed.

Despite assumptions and simplifications, the linear distribution of $C_f/C_h$ on the windward of curved surfaces is revealed by theoretical analyses and shown by DSMC simulations with different types of shapes. For circular cylinder cases, with the available explicit $f''_w/g'_w$ and $\hat{p}_0$, good agreements are observed between the present analytical prediction and numerical results.

\section{Ratio of Skin Friction to Heat Transfer in Rarefied Flows}

The analyses in the above section are based on the boundary layer assumption, and thus the theory is valid only for continuous flows. When the flow deviates from the continuum regime, the nonequilibrium of molecular collisions causes non-linear shear and heat transfer, and then the linear Newtonian shear and Fourier heat transfer in the N-S equations fail. If the deviation is small, the constitutive relation can be corrected by bringing in the second order shear and heat transfer in Burnett equations \cite{chapman1970} as has been suggested by Wang et al. \cite{wang2009,wang2010}. In order to obtain a more general analogy relation covering the rarefied flows, the second order shear stress and heat transfer will be studied firstly in the near continuum regime, and then, based on numerical validation and calibration, the results will be extended to the more rarefied flow regime to calculate $C_f/C_h$. In fact, an early exploration has been carried out in our previous work on the flat plates leading edge flow problem \cite{chen2014}.

Instead of directly solving the Burnett Equations, the second order effects could be evaluated based on a perturbation point of view, i.e. to use the non-linear constitutive relations in Burnett equations to analyze the flow field features predicted by the first order approximation for instance the boundary layer theory or the computation method of N-S equations.
The original form of the Burnett equations can be found in Chapman and Cowling's derivations \cite{chapman1970}. For planar bodies, taking the assumptions $\partial p/\partial y\approx0$, $v|_w=0$, $\frac{\partial u}{R\partial \theta}\big|_w=-\frac{\partial v}{\partial y}\big|_w\approx\sqrt{\frac{\gamma-1}{\gamma}}\frac{TU_\infty}{T_0R}\big|_w $ \cite{matting1964} and the constant wall temperature $T|_w=T_w$. Besides, in the near continuum regime we have $\frac{\partial u}{\partial y}\big|_w\gg{u_w}/{R}$ and $\frac{\partial T}{T\partial y}\big|_w\gg{1}/{R}$, then the Burnett shear and heat transfer near a planar curved surface become:
\begin{equation}
  \begin{split}
    \tau^{(2)}_{w} &\approx \omega_3\frac{\mu^2}{\rho RT}\frac{\partial^2T}{\partial \theta \partial y}\bigg|_w+\frac{1}{2}\omega_4\frac{\mu^2}{\rho RpT}\frac{\partial p}{\partial \theta}\frac{\partial T}{\partial y}\bigg|_w-\omega_2\frac{\mu^2}{p}\frac{\partial u}{R\partial \theta}\frac{\partial u}{\partial y} \bigg|_w
      \\
    \dot{q}_w^{(2)}&\approx\frac{1}{2}\theta_3\frac{\mu^2}{\rho Rp}\frac{\partial p}{\partial \theta}\frac{\partial u}{\partial y}\bigg|_w-\frac{1}{2}\theta_4\frac{\mu^2}{\rho R}\frac{\partial^2u}{\partial \theta\partial y}\bigg|_w+(2\theta_2-3\theta_5)\frac{\mu^2}{\rho TR}\frac{\partial T}{\partial y}\frac{\partial u}{\partial \theta} \bigg|_w
  \end{split}\label{eq_burnett}
\end{equation}
As has been testified in flows past the leading edge of flat plates, the above simplifications, although not strictly, still reflect the essential features of non-linear terms \cite{chen2014}.

The flow field predicted by the first order continuum theory and the gradients near the wall need to be given before using Eq. (\ref{eq_burnett}) for further discussions. Assuming the normalized pressure and heat transfer distribution functions as
\begin{equation}\label{eq_distri}
	\tilde{p}(\theta)=p/p_0~~~~~~\varsigma(\theta)=\dot{q}_w^{(1)}/\dot{q}_{w,0}^{(1)}
\end{equation}
where $p_0$ and $\dot{q}_{w,0}$ are the pressure and the heat flux at the stagnation point, respectively. Both $\tilde{p}$ and $\varsigma$ are functions of $\theta$ and are able to be obtained with different approaches such as analytical theories or data correlations \cite{lees1956,kemp1959,murzinov1966,beckwith1961}.
Taking $p_0\approx\gamma M_{\infty}^2p_{\infty}$ in hypersonic limit, then Eq. (\ref{eq_distri}) can be transformed to
\begin{equation}\label{eq_p_parT}
	p\approx\gamma M_{\infty}^2p_{\infty}\tilde{p}~~~~~~\frac{\partial T}{\partial y}\bigg|_w=\frac{\dot{q}^{(1)}_{w,0}}{k_w}\varsigma
\end{equation}
And with the linear analogy equation (\ref{eq_linear_the}), the normal gradient of $u$ becomes
\begin{equation}\label{eq_part_u}
	\frac{\partial u}{\partial y}\bigg|_w=\left(\frac{k}{\mu}\frac{\partial T}{\partial y}\right)_w\frac{C_r}{U_{\infty}}\theta
\end{equation}
With Eqs.~(\ref{eq_p_parT}) and  (\ref{eq_part_u}) submitted, Eq. (\ref{eq_burnett}) becomes:
\begin{equation}\label{eq_burnett_2}
	\begin{split}
	\tau^{(2)}_{w}&\approx\frac{\omega_3\varsigma_{\theta}/\varsigma+0.5\omega_4{\tilde{p}}_{\theta}/\tilde{p}}{\theta C_r}\left(\frac{\mu^2U_{\infty}}{k\rho RT}\mu\frac{\partial u}{\partial y} \right)_w-2\omega_2\sqrt{\gamma/(\gamma-1)}\left(\frac{\mu^2}{\rho U_\infty R}\frac{\partial u}{\partial y} \right)_w
	\\
	\dot{q}_w^{(2)}&\approx\tfrac{1}{2} \big[\theta_3C_r(\theta{\tilde{p}_{\theta}}/\tilde{p})-\theta_4C_r(1+\theta\varsigma_{\theta}/\varsigma)+4Pr(2\theta_2-3\theta_5)\sqrt{(\gamma-1)/\gamma}\big]\left(\frac{\mu}{\rho RU_{\infty}}k\frac{\partial T}{\partial y}\right)_w
	\end{split}
\end{equation}
where $\varsigma_{\theta}=d \varsigma/d \theta$, $\tilde{p}_{\theta}=d \tilde{p}/d \theta$. In Eq. (\ref{eq_burnett_2}), expressions of the second order shear stress and heat transfer contain $\mu\frac{\partial u}{\partial y}$ and $k\frac{\partial T}{\partial y}$ explicitly, and thus, more meaningfully, we could get the relative magnitudes of the second order effects
\begin{equation}
	\begin{split}\label{eq_burnett_rate}
		\frac{\tau_w^{(2)}}{\tau_w^{(1)}}\bigg|_r&\approx\frac{\omega_3\varsigma_{\theta}/\varsigma+0.5\omega_4{\tilde{p}_{\theta}}/\tilde{p}}{\theta C_r}\frac{\mu_r^2U_{\infty}}{k_r\rho_r RT_r}-2\omega_2\sqrt{\gamma/(\gamma-1)}\frac{\mu_r}{\rho_r U_\infty R}
		\\
		\frac{\dot{q}_w^{(2)}}{\dot{q}_w^{(1)}}\bigg|_r&\approx \tfrac{1}{2} \big[\theta_4C_r(1+\theta\varsigma_{\theta}/\varsigma)-\theta_3C_r(\theta{\tilde{p}_{\theta}}/\tilde{p})-4Pr(2\theta_2-3\theta_5)\sqrt{(\gamma-1)/\gamma}\big]\frac{\mu_r}{\rho_rR U_{\infty}}
	\end{split}
\end{equation}
The subscript $r$ represents the quantities under the reference temperature $T_r$, as in hypersonic flows the reference temperature method is usually employed to present characteristics of boundary layers. Here $T_r=\frac{1}{2}(T_w+T_0)$, and the viscosity-temperature power law $\mu_r/\mu_{\infty}=(T_r/T_{\infty})^{\omega}$ is applied. Then Eq. (\ref{eq_burnett_rate}) becomes
\begin{equation}\label{eq_wr}
	\begin{split}
	\frac{\tau_w^{(2)}}{\tau_w^{(1)}}\bigg|_r&=\Gamma_1(\theta)W_r
	\\
	\frac{\dot{q}_w^{(2)}}{\dot{q}_w^{(1)}}\bigg|_r&=\Gamma_2(\theta)W_r
	\end{split}
\end{equation}
where $W_r=M_{\infty}^{2\omega}/Re_l$, $Re_l=\rho_{\infty}RU_{\infty}/\mu_{\infty}$, $\Gamma_1$ and $\Gamma_2$ are:
\begin{equation} \label{eq_Gamma}
	\begin{split}
	\Gamma_1&=\frac{Pr(\gamma-1)^{1+\omega}}{\gamma\tilde{p}}\frac{\omega_3\varsigma_{\theta}/\varsigma+0.5\omega_4{\tilde{p}}_{\theta}/\tilde{p}}{\theta C_r}\left(\frac{T_w+T_0}{4T_0} \right)^{\omega}-2\omega_2\frac{(\gamma-1)^{\omega+1/2}}{\gamma^{1/2}}\left(\frac{T_w+T_0}{4T_0} \right)^{\omega+1}
	\\
	\Gamma_2&=\frac{(\gamma-1)^{1+\omega}}{2\gamma\tilde{p}}\big[\theta_4C_r(1+\tfrac{\theta\varsigma_{\theta}}{\varsigma})-\theta_3C_r\tfrac{\theta{\tilde{p}}_{\theta}}{\tilde{p}}-4Pr(2\theta_2-3\theta_5)\sqrt{(\gamma-1)/\gamma}\big]\left(\frac{T_w+T_0}{4T_0} \right)^{\omega+1}
	\end{split}
\end{equation}
With $\tau_w^{(2)}$ and $\dot{q}_w^{(2)}$ being added to calculate the skin friction and heat transfer, we can introduce a modification to the linear analogy Eq. (\ref{eq_linear_the}):
\begin{equation}\label{eq_ratio_mod}
	\frac{C_f}{C_h}=U_{\infty}\frac{\tau_w^{(1)}+\tau_w^{(2)}}{\dot{q}_w^{(1)}+\dot{q}_w^{(2)}}=\frac{1+\Gamma_1W_r}{1+\Gamma_2W_r}C_r\theta
\end{equation}
Values of $\Gamma_1$ and $\Gamma_2$ denote the influence of the body shape and vary with $\theta$, and generally both of them are in the order of unit.
For a typical case of the hypersonic nitrogen gas flows past a circular cylinder under $T_w/T_0\approx0$, with the modified Newtonian pressure and the heat transfer fitting formula from Beckwith and Gallagher \cite{beckwith1961} accepted in Eq. (\ref{eq_Gamma}), an approximation can be taken in the range of $0\le\theta\le \pi/6$ as \[\Gamma_1-\Gamma_2\approx0.43+0.35\theta^2+0.18\theta^4\]
from the Taylor series expansion. Then in the near continuum regime with $W_r\ll1$, Eq. (\ref{eq_ratio_mod}) can be simplified to
\begin{equation}\label{eq_ratio_wr}
	\frac{C_f}{C_h}\approx[1+(\Gamma_1-\Gamma_2)W_r]C_r\theta
\end{equation}

\begin{figure}[htbp]
	\centering
	\begin{minipage}[t]{0.5\linewidth}
	\centering
	\begin{overpic}[width=200pt]{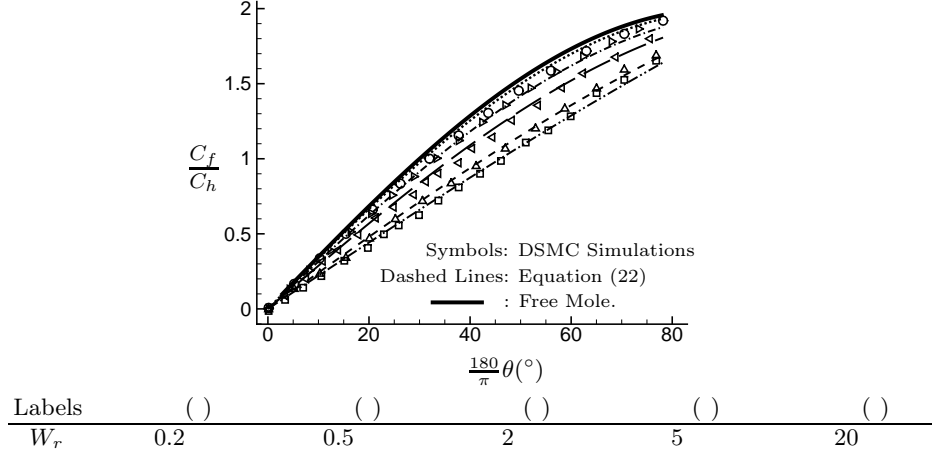}
	\put(53,0){$\tfrac{180}{\pi}\theta(^{\circ})$}
	\large
	\put(0,38){$\frac{C_f}{C_h}$}
	\scriptsize
	\put(45.3,23){Symbols: DSMC Simulations}
	\put(36.6,18){Dashed Lines: Equation (\ref{eq_bridge_cylinder})}
	\put(60,13.3){: Free Mole.}
	\put(-35,-10){
	\small
	\begin{tabular}
	{p{25pt}<{\centering}p{60pt}<{\centering}p{60pt}<{\centering}p{60pt}<{\centering}p{60pt}<{\centering}p{60pt}<{\centering}}
	Labels&
      \tikz[baseline=-2pt,dash pattern=on 8 off 1.3 on 0.8 off 1.3 on 0.8 off 1.3]
      \draw[line width=0.8] (0,0)--(18,0);
      (\tikz[baseline=-1pt] \draw[line width=0.7pt,scale=0.7] (0,0) rectangle (3.5,3.5);)
      &
      \tikz[baseline=-2pt,dash pattern=on 4 off 4]
      \draw[line width=0.8] (0,0)--(18,0);
      (\tikz[baseline=-1pt] \draw[line width=0.7pt,scale=0.7] (0,0)--(2,4)--(4,0)--cycle;)
      &
      \tikz[baseline=-2pt,dash pattern=on 12 off 2]
      \draw[line width=0.8] (0,0)--(18,0);
      (\tikz[baseline=-1pt] \draw[line width=0.7pt,scale=0.7,rotate=90] (0,0)--(2,4)--(4,0)--cycle;)
      &
      \tikz[baseline=-2pt,dash pattern=on 5 off 1.3 on 0.8 off 1.3]
      \draw[line width=0.8] (0,0)--(18,0);
      (\tikz[baseline=-3pt] \draw[line width=0.7pt,scale=0.7,rotate=270] (0,0)--(2,4)--(4,0)--cycle;)
      &
      \tikz[baseline=-2pt,dash pattern=on 1 off 1]
      \draw[line width=0.8] (0,0)--(18,0);
      (\tikz[baseline=-1pt] \draw[line width=0.7pt,scale=0.7] (2,2) circle(2);)
      \\
      \hline
      $W_r$&$0.2$&$0.5$&$2$&$5$&$20$\\
	\end{tabular}
	}
	\end{overpic}
    \vspace{30pt}
	\end{minipage}
	\caption{Non-linear analogy in hypersonic rarefied flows past circular cylinders with $M_{\infty}=10\sim25$, $T_w/T_{0}=0.01$.}\label{fig_bridge_circle}
\end{figure}

The first order correction $(\Gamma_1-\Gamma_2)W_r$ is similar with Wang et al.'s result in the stagnation point heat transfer problem \cite{wang2010}. It turns out that, $W_r$ is a control parameter of the rarefied gas effects not only at the stagnation point but also in the downstream region.
Although the correction factor in Eq. (\ref{eq_ratio_wr}) is explicit and clear, it will lose its credibility when the rarefaction degree of the flow is sufficiently high, and due to the complexity of the transition flow, the data fitting and calibrating seem still unavoidable to get a practical general analogy. In fact, based on the rarefied flow criterion $W_r$, a bridge function can be built between the continuum limit Eq. (\ref{eq_linear_the}) and the free molecular limit $\large(C_f/C_h\large)_{free}\approx2sin\theta$.
For circular cylinders, the function is carried out as
\begin{equation}\label{eq_bridge_cylinder}
	\frac{C_f}{C_h}=\frac{C_r\theta+1.4sin\theta W_r}{1+0.7W_r}
\end{equation}
Eq. (\ref{eq_bridge_cylinder}) with different $W_r$ is plotted in Fig. (\ref{fig_bridge_circle}) compared with the DSMC simulations of nitrogen gas flows past circular cylinders. It can be seen that the variation of $C_f/C_h$ with $\theta$ is no longer linear in the rarefied gas flow regime.

\section{Conclusions and Remarks}
The relation between skin friction and heat transfer for blunt-nosed bodies in hypersonic flows, named the general Reynolds analogy, has been investigated in this paper by using the theoretical modelling and the DSMC methods.

First, based on the boundary layer flow properties, the ratio of the skin friction to the heat transfer for the blunt-nosed body was found proportional to the local surface slope angle. As a typical demonstration, an explicit expression of the ratio was derived for circular cylinders. Also, numerical calculations indicated that this characteristic exists for other blunt-nosed shapes even in chemically reactive flows.

Second, the analogy in rarefied gas flows was analyzed. In the rarefied flow regime, the  deviation from the linear distribution of the ratio was proved to be controlled by the rarefied flow criterion $W_r$. Therefore, a bridge function was constructed based on $W_r$ to describe the analogy in the transition flow regime.

This study, combined with our former investigation on the flat plate leading edge flows \cite{chen2014}, clarifies the general Reynolds analogy in the whole flow regime for both flat plates and blunt-nosed bodies in hypersonic flows.

The present general Reynolds analogy has potential usefulness in the engineering practice. From the analogy relation, the skin friction is related to the heat flux along surfaces, or further to the heat flux at the stagnation point and its normalized distribution downstream. As a result, the viscous drag integrated from the skin friction is proportional to the stagnation point heat flux, which suggests that if one of them is known, the other could also be obtained immediately.

\section*{Acknowledgments}
This work was supported by the National Natural Science Foundation of China (Grant No. 11202224). The paper benefits a lot from our discussions with Associate Professor Lin Bao.


\section*{References}
\begin{thebibliography}{}
\bibitem{lees1956} Lees, L., ``Laminar Heat Transfer over Blunt-nosed Bodies at Hypersonic Flight Speeds,'' \textit{Journal of Jet Propulsion}, Vol. 26, No. 4, 1956, pp. 259--269.
\bibitem{fay1958} Fay, J. A., and Riddell, F. R., ``Theory of Stagnation Point Heat Transfer in Dissociated Air,'' \textit{Journal of the Aerospace Sciences}, Vol. 25, No. 2, 1958, pp. 73--85.
\bibitem{kemp1959} Kemp, N. H., and Rose, P. H. and Detra, R. W., ``Laminar Heat Transfer Around Blunt Bodies in Dissociated Air,'' \textit{Journal of the Aerospace Sciences}, Vol. 26, No. 7, 1959, pp. 421--430.
\bibitem{murzinov1966} Murzinov, I. N., ``Laminar Boundary Layer on a Sphere in Hypersonic Flow of Equilibrium Dissociating Air,'' \textit{Fluid Dynamics}, Vol. 1, No. 2, 1966, pp. 131--133.
\bibitem{beckwith1961} Beckwith, I. E., and Gallagher, J. J., ``Local Heat Transfer and Recovery Temperatures on a Yawed Cylinder at a Mach Number of 4.15 and High Reynolds Numbers,'' NASA TR-R104, May 1962.
\bibitem{cheng1961} Cheng, H. K., ``Hypersonic Shock-Layer Theory of the Stagnation Region at Low Reynolds Number,'' \textit{Proceedings of the 1961 Heat Transfer and Fluid Mechanics Institute}, edited by Binder, R. C., and Epstein, M., and Mannes, R. L. and Yang, H. T., Stanford University Press, Chicago, 1961, pp. 161--175.
\bibitem{wang2009} Wang, Z. H., Bao, L., and Tong, B. G.,``Variation Character of Stagnation Point Heat Flux for Hypersonic Pointed Bodies from Continuum to Rarefied Flow States and Its Bridge Function Study,'' \textit{Science in China Series G: Physics, Mechanics and Astronomy}, Vol. 52, No. 12, 2009, pp. 2007-2015.
\bibitem{wang2010} Wang, Z. H., Bao, L., and Tong, B. G.,``Rarefaction Criterion and Non-
Fourier Heat Transfer in Hypersonic Rarefied Flows,'' \textit{Physics of Fluids}, Vol. 22, No. 12, 2010, Paper 126103.
\bibitem{santos2002} Santos, W. F., and Lewis, M. J., ``Power-law Shaped Leading Edges in Rarefied Hypersonic Flow,'' \textit{Journal of Spacecraft and Rockets}, Vol. 39, No. 6, 2002, pp. 917--925.
\bibitem{bird1994} Bird, G. A., \textit{Molecular Gas Dynamics and the Direct Simulation of Gas Flows}, Oxford Univ. Press, New York, 1994.
\bibitem{anderson2006} Anderson, J. D., \textit{Hypersonic and High Temperature Gas Dynamics},
2nd ed., McGraw–Hill, New York, 2006, Chaps. 2, 6.
\bibitem{lees1955} Lees, L., ``Hypersonic Flow,'' \textit{Journal of Spacecraft and Rockets}, Vol. 40, No. 5, 1955, pp. 700--735.
\bibitem{korobkin1954} Korobkin,I., ``Laminar Heat Transfer Characteristics of a Hemisphere for the Mach Number Range 1.9 to 4.9,'' U. S. Naval Ordnance Laboratory, NAVORD Report no. 3841, October 10, 1954.
\bibitem{cohen1956} Cohen, C. B., and Reshotko, E., ``Similar Solutions for the Compressible Laminar Boundary Layer with Heat Transfer and Pressure Gradient,'' NACA Report 1293, 1956.
\bibitem{blottner1970} Blottner, F. G., ``Finite Difference Methods of Solutions of the Boundary-layer Equations,'' \textit{AIAA Journal}, Vol. 8, No. 2, 1970, pp. 193--205.
\bibitem{boyd2011} Holman, T. D. and Boyd, I. D., ``Effects of Continuum Breakdown on Hypersonic Aerothermodynamics for Reacting Flow,'' \textit{Physics of Fluids}, Vol. 23, No. 2, 2011, Paper 027101.
\bibitem{chapman1970} Chapman, S., and Cowling, T. G., \textit{The Mathematical Theory of Non-Uniform Gases}, 3rd ed., Cambridge Univ. Press, Cambridge, England, U.K., 1970, pp. 280--296.
\bibitem{chen2014} Chen, X. X., Wang, Z. H., and Yu, Y. L., ``Nonlinear Shear and Heat Transfer in Hypersonic Rarefied Flows Past Flat Plates,'' \textit{AIAA Journal}, (2013), accessed April 4, 2014. doi: 10.2514/1.J053168
\bibitem{matting1964} Matting, F. W., ``General Solution of the Laminar Compressible Boundary Layer in the Stagnation Region of Blunt Bodies in Axisymmetric Flow,'' NASA Technical Note, D-2234, 1964.
\end{thebibliography}
\end{document}